\documentclass[prl,showpacs,amsmath,amssymb,twocolumn]{revtex4}
\usepackage{graphicx}
\usepackage{array}
\usepackage{dcolumn}
\usepackage{bm}
\usepackage{hyperref}

\newcommand{\figurewidth}{\columnwidth}

\begin{document}

\title{Storage-ring measurement of the hyperfine induced
$^\mathbf{47}$Ti$^\mathbf{18+}$($\mathbf{2s\,2p\;^3P_0 \to 2s^2\;^1S_0}$) transition rate}

 \author{S. Schippers}
 \author{E. W. Schmidt}
 \author{D. Bernhardt}
 \author{D. Yu}\altaffiliation[Permanent address: ]{Institute of Modern Physics,
           Chinese Academy of Sciences, Lanzhou 730000, P. R. China}
 \author{A. M{\"u}ller}
 \affiliation{Institut f{\"u}r Atom- und Molek{\"u}lphysik, Justus-Liebig-Universit{\"a}t, 35392
 Giessen, Germany}

 \author{M. Lestinsky}
 \author{D. A. Orlov}
 \author{M. Grieser}
 \author{R. Repnow}
 \author{A. Wolf}
 \affiliation{Max-Planck-Institut f{\"u}r Kernphysik, 69117 Heidelberg, Germany}

\date{\today}

\begin{abstract}
The hyperfine induced $2s\,2p\;^3P_0 \to 2s^2\;^1S_0$ transition rate
$A_\mathrm{HFI}$ in berylliumlike $^{47}$Ti$^{18+}$ was measured.
Resonant electron-ion recombination in a heavy-ion storage ring was
employed to monitor the time dependent population of the $^3P_0$ state.
The experimental value $A_\mathrm{HFI}=0.56(3)$~s$^{-1}$ is almost 60\%
larger than theoretically predicted.
\end{abstract}



\maketitle

\begin{figure}[ttt]
\includegraphics[width=0.9\figurewidth]{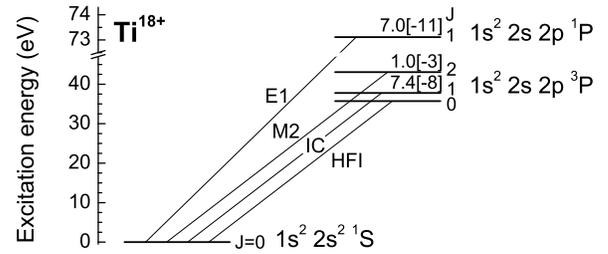}
\caption{\label{fig:Ti18levels} Simplified level diagram for
berylliumlike Ti$^{18+}$. The level energies were taken from the NIST
Atomic Spectra Data Base \cite{Ralchenko2005a}, and the lifetimes (in s)
labeling the excited levels were calculated from theoretical one-photon
transition rates \cite{Bhatia1980a} that do not account for hyperfine
effects. Numbers in square brackets denote powers of 10. In case of
nonzero nuclear spin the hyperfine induced (HFI) $^3P_0 \to {^1S_0}$ transition
rate $A_\mathrm{HFI}$ acquires a finite value. The $^1P_1 \to {^1S_0}$ electric dipole (E1), $^3P_2 \to {^1S_0}$ magnetic quadrupole (M2), and $^3P_1 \to {^1S_0}$ intercombination (IC) transitions are also shown.}
\end{figure}

Atoms and ions in metastable excited states with very small
electromagnetic transition rates are promising systems for realizing
ultraprecise atomic clocks, for the diagnostic of astrophysical media
regarding the competition of radiative and non-radiative processes, for
realizing novel types of cold atomic gases, and for probing fundamental
correlation effects in the bound states of few-electron systems.  In
particular, in alkaline-earth-like and, in general, divalent atoms and
ions, having a $(ns)^2\;^1S_0$ ground state and a valence shell $n$, the
first excited level above the ground state is the term $ns\,np\;^3P_0$
(Fig.\ \ref{fig:Ti18levels}). The absence of a total electronic angular
momentum $J$ for this level makes its single photon decay to the ground
state impossible except for the hyperfine induced decay in the case of a
nucleus with a spin $I\ne0$. For $I\ne0$ the hyperfine interaction mixes
states with different $J$ and the $ns\,np\;^3P_0$ term aquires a finite,
but long and strongly isotope-dependent radiative lifetime.  These
hyperfine-dominated decay rates have been treated theoretically for
beryllium-, magnesium-, and zinclike ions
\cite{Marques1993,Marques1993a,Brage1998a,Liu2006a} and for divalent
heavier atoms \cite{Porsev2004a,Santra2004a}, where the long and
isotope-dependent lifetimes are attractive in view of obtaining
ultraprecise optical frequency standards and for cold-atom studies.  In
low-density astrophysical systems, the fluorescence observed from the
hyperfine-induced radiative decay of the long-lived $2s2p\;^3P_0$ level in
the berylliumlike ion $^{13}$C$^{2+}$ can be used to infer the
$^{13}$C/$^{12}$C abundance ratio, giving insight into stellar
nucleosynthesis \cite{Rubin2004a}.

Hyperfine-induced (HFI) decay rates in divalent $ns\,np\;^3P_0$ states
were so far determined experimentally only for the atomic-clock
transition $5s5p\;^3P_0$-$5s^2\;^1S_0$ of $^{115}$In$^+$ in a
radio-frequency ion trap \cite{Becker2001a}, with an $\sim$5\%
uncertainty, and for the beryllium-like ion N$^{3+}$ \cite{Brage2002a}
using observations from a planetary nebula and yielding an uncertainty of
33\%.  The $^{115}$In$^+$ result is in good agreement with estimates on
hyperfine mixing based on experimental spectral parameters of the atom
\cite{Becker2001a}, while the N$^{3+}$ result, even at its limited
precision, allows one to discriminate between the lifetimes predicted by
atomic structure calculations \cite{Marques1993,Brage1998a} that differ
by a factor of almost 4.  In view of the large theoretical discrepancies
in the atomic-structure based calculations and the scarce experimental
data, accurate experimental benchmarks are highly desirable, especially
for few-electron systems such as beryllium-like ions with particularly
strong correlation effects.

Among the few-electron systems, HFI transitions were also studied for
highly charged helium-like ions \cite{Gould1974a}.  However, in contrast
to the present case they do not represent the only radiative decay path,
but compete with allowed transitions such as $1s2p\;^3P_0$-$1s2s\;^3S_1$.
Moreover, the resulting lifetimes are generally much shorter than in a
divalent system with comparable valence shell and nuclear charge; the HFI
decay rate variations studied so far for the $1s2p\;^3P_0$-$1s^2\;^1S_0$
line are in the range of $10^7$--$10^{12}$~ s$^{-1}$
\cite{Engstroem1981a,Dunford1991,Indelicato1992,Birkett1993a}.

In the present work we obtain an experimental value for the decay rate of
the long-lived $2s\,2p\;^3P_0$ state in $^{47}$Ti$^{18+}$ ($I=5/2$)
through radiative transitions induced by the hyperfine interaction only.
This decay limits the lifetime of the $2s\,2p\;^3P_0$ state in
$^{47}$Ti$^{18+}$ to $\sim$1.8~s.  A benchmark HFI decay rate for this
highly charged beryllium-like system is obtained with an uncertainty of
$\sim$5\% (almost an order of magnitude lower than that accomplished
previously \cite{Brage2002a}); the observed decay rate is significantly
larger than the only available theoretical prediction \cite{Marques1993},
which instead would predict a lifetime of $\sim$2.8~s.  The
experiment uses fast, isotopically pure ion beams of $^{47}$Ti$^{18+}$
and $^{48}$Ti$^{18+}$ ($I=0$) circulating for up to 200~s in the heavy-ion
storage ring TSR of the Max-Planck Institute for Nuclear Physics,
Heidelberg, Germany.  The new high-resolution electron-ion collision
spectrometer \cite{Sprenger2004a} at this facility is used to detect a
signal proportional to Ti$^{18+}$ ions in the metastable $2s\,2p\;^3P_0$
state.  For this purpose, the electron-ion collision energy is tuned to a
value where dielectronic recombination (DR) occurs only for ions in this
excited level, and the rate of recombined Ti$^{17+}$ ions produced at
this collision energy is recorded as a function of the storage time. This
method had been applied previously for measuring the slow radiative decay
rates of $1s\,2s\;^3S$ states in the He-like ions (B$^{3+}$, C$^{4+}$,
N$^{5+}$ \cite{Schmidt1994} and Li$^{+}$ \cite{Saghiri1999}).

Mass selected $^{47,48}$Ti$^{18+}$ ion beams (natural abundances 7.2\%
and 73.7\%, respectively) were provided by a tandem accelerator, followed
by a radio-frequency linear accelerator, at energies close to 240 MeV,
using a fixed magnetic setting for the beam line and the storage ring
(magnetic rigidity 0.8533~Tm).  The residual gas pressure in the storage
ring was $<5\times10^{-11}$~mbar.  In one straight section of the storage
ring (circumference $C=55.4$~m) the ion beam was continuously phase-space
cooled using the velocity-matched electron beam of the TSR electron
cooler (electron density $\sim 5.0\times10^{7}$ cm$^{-3}$).  In a second
straight section, the ion beam was merged with the collinear electron
beam of the high-resolution electron target \cite{Sprenger2004a}, run at
variable acceleration voltage in order to set the required collision
energy in the co-moving reference frame of the ions.  At collision
energies of 0--2 eV, the electron density in the electron target was
$5.6\times10^{7}$ cm$^{-3}$.  Ti$^{17+}$ ions formed by electron-ion
recombination in the electron target or by charge transfer in collisions
with residual gas molecules were deflected out of the closed orbit of the
circulating Ti$^{18+}$ ion beam in the first dipole magnet downstream of
the electron target and were directed onto a scintillation detector
operated in single-particle counting mode with nearly 100\% detection
efficiency and negligible dark count rate.  The overlap lengths were
$\sim$1.5 m each in both interaction regions.  The cooled ion beam
velocities for the two isotopes, as obtained from the space-charge
corrected \cite{Kieslich2004a} electron acceleration voltage at velocity
matching, were $\beta^\mathrm{(48)}=0.1026(1)$ and
$\beta^\mathrm{(47)}=0.1047(1)$ (in units of the vacuum speed of light).

\begin{figure}[ttt]
\includegraphics[width=\figurewidth]{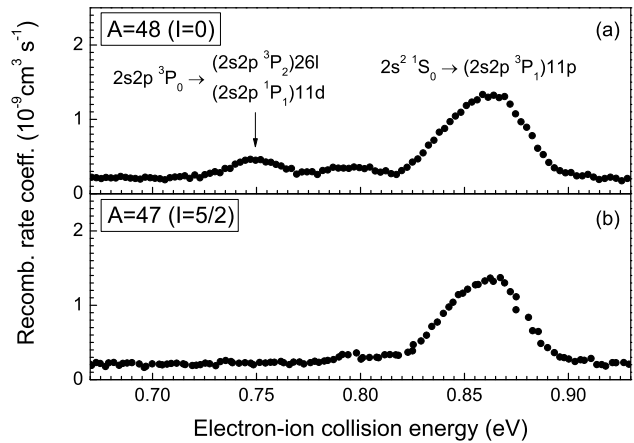}
\caption{\label{fig:DR} Measured rate coefficients for the
electron-ion recombination of $^{48}$Ti$^{18+}$ (upper panel) and
$^{47}$Ti$^{18+}$ (lower panel) in the energy range of interest. The resonances that are formed by resonant dielectronic capture of the initially free electron are assigned to
doubly excited Ti$^{17+}$ states as indicated with the aid of theoretical calculations using
the \textsc{autostructure} code \cite{Badnell1986}. A comprehensive plot of the $^{48}$Ti$^{18+}$ recombination spectrum extending over the entire experimental energy range of 0--80~eV can be found in Ref.\ \cite{Schippers2006b}.}
\end{figure}

Recombination spectra of the Ti$^{18+}$ ions as a function of the
relative electron-ion energy, $E_\mathrm{rel}$, were taken by varying the
cathode voltage of the electron target appropriately.  The procedure for
electron-ion measurements at the TSR storage ring has been described in
more detail in, e.\,g., Ref.\ \cite{Kieslich2004a} (and references
therein).  For the present spectral measurements, a constant current of
cooled, circulating Ti$^{18+}$ ions was maintained.  Currents of a
fraction of a $\mu$A were injected at a rate of $\sim$1 s$^{-1}$ in order
to obtain stationary stored currents of $\sim$40~$\mu$A for
$^{48}$Ti$^{18+}$ and $\sim$6~$\mu$A for $^{47}$Ti$^{18+}$, respectively;
this largely reflects the difference in the natural isotope abundances.
The average storage lifetime in this mode is $\sim$50 s.

Figure \ref{fig:DR} shows a region of the recombination spectrum where
resonances of metastable Ti$^{18+}$($2s2p\;^3P_0$) ions occur close to a
strong resonance from ground-state Ti$^{18+}$ ions. Theoretical
calculations using the \textsc{autostructure} code \cite{Badnell1986}
were performed to assign the weaker structures to the metastables, which
decay by collisional interactions for only the $^{48}$Ti$^{18+}$ beam
[Fig.\ \ref{fig:DR}(a)]. Using a $^{47}$Ti$^{18+}$ beam [Fig.\
\ref{fig:DR}(b)], the resonances assigned to metastable
Ti$^{18+}$($2s2p\;^3P_0$) ions essentially disappear, as their average
population is strongly reduced through the radiative decay (expected life
time 2.8 s \cite{Marques1993}). The isotope-dependent occurrence of DR
resonances was also observed in earlier TSR experiments, using the
heavier divalent ion Pt$^{48+}$ (Zn-like) \cite{Schippers2005b}. The
\textsc{autostructure} calculations of the recombination spectrum
indicate an average population of $\sim$5\% for the metastable
Ti$^{18+}$($2s2p\;^3P_0$) ions in the stored $^{48}$Ti$^{18+}$ beam. A
similar average population of the $2s2p\;^3P_0$ metastable state was
found in a recent DR experiment with Be-like $^{56}$Fe$^{22+}$
\cite{Savin2006a}.

For the determination of the time constant associated with the hyperfine
quenching the decay of the $^{47}$Ti$^{18+}$($^3P_0$) beam component was
monitored as a function of storage time. To this end the relative
electron-ion energy in the electron target was set fixed to 0.75~eV where
a DR resonance associated with excitation of the $^3P_0$ state occurs
(vertical arrow in Fig.\ \ref{fig:DR}). After injection of a single
Ti$^{18+}$ ion pulse into the storage ring the recombination rate was
recorded for up to 200~s. Prior to the injection of the next pulse the
remaining ions were kicked out of the ring. This scheme was repeated to
reduce statistical uncertainties to a suitable level.

\begin{figure}[ttt]
\includegraphics[width=\figurewidth]{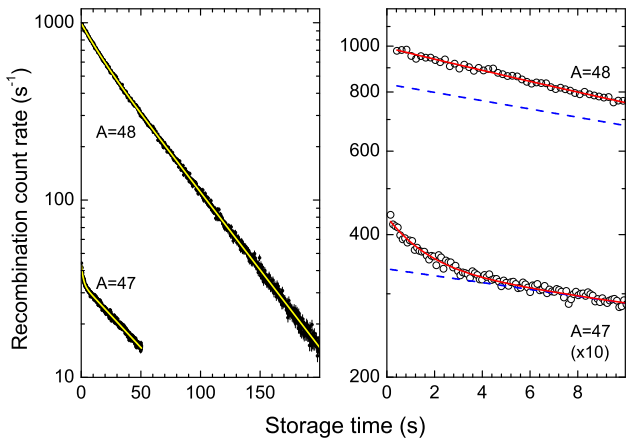}
\caption{\label{fig:life} Measured $^{48}$Ti$^{17+}$  and
$^{47}$Ti$^{17+}$ recombination count rates as a function of storage
time. Both panels show the same curves on different time scales. Clearly,
the $A=47$ curve has a fast decaying component that is absent in the
$A=48$ curve. The solid lines are the results of the fits of Eq.\
\ref{eq:F} to the experimental data points (symbols). In the right panel
the fitted $c_g^\mathrm{(A)}\exp(-\lambda^\mathrm{(A)}_gt)$ components
are shown as dashed lines. The fit results are summarized in Tab.
\protect\ref{tab:fit}.}
\end{figure}

Figure \ref{fig:life} displays the two decay curves that were
obtained for the two isotopes with $A=48$ and $A=47$. Since the
recombination signal was produced by both $^1S_0$ and $^3P_0$ ions
the sum of two exponentials, i.\,e., the function
\begin{equation}\label{eq:F}
  F^\mathrm{(A)}(t) = c^\mathrm{(A)}_{m} e^{-\lambda^\mathrm{(A)}_mt}+ c^\mathrm{(A)}_{g}e^{-\lambda^\mathrm{(A)}_gt}
\end{equation}
was fitted to the measured decay curves \cite{Schmidt1994,Saghiri1999}.
As discussed in more detail below, the $^1S_0$ state contributes also at
$E_\mathrm{rel}=0.75$~eV to the measured recombination signal by
nonresonant radiative recombination and by electron capture from the
residual gas. The fit results are listed in Tab.\ \ref{tab:fit}. The
one-sigma confidence limits (numbers in parentheses) on the fit
parameters $c^\mathrm{(A)}_m$, $\lambda^\mathrm{(A)}_m$,
$c^\mathrm{(A)}_{g}$ , and $\lambda^\mathrm{(A)}_g$ were obtained by
Monte-Carlo simulations \cite{Press2002a} of 100 synthetic data sets for
each isotope.

The interpretation of the fitted decay constants in terms of atomic
transition rates is straightforward if the rate for collisional
excitation of the $^3P_0$ state during collisions of the Ti$^{18+}$ ions
with residual gas particles is negligibly small. Collisional excitation
has been investigated in previous experimental measurements of metastable
state lifetimes, e.\,g., with stored Xe$^+$ \cite{Mannervik1997a},
C$^{4+}$ \cite{Schmidt1994} and Li$^+$ \cite{Saghiri1999} ions. In all
these cases collisional processes were found to have a negligibly small
influence on the population of the metastable states under investigation,
mainly because of the low residual particle density in the ultrahigh
vacuum of the storage ring. Since the residual gas pressure and the
collisional excitation cross sections of the present experiment and of
the above mentioned studies are of the same order of magnitude,
collisional excitation and deexcitation processes are assumed to be
negligible within the experimental uncertainty.

\begin{table}[ttt]
\caption{\label{tab:fit}Results (including statistical uncertainties) for the decay constants
$\lambda^\mathrm{(A)}_{m,g}$ and relative weights $c^\mathrm{(A)}_{m,g}$ obtained from the fits of
Eq.\ \ref{eq:F} to the experimental decay curves (Fig.\ \ref{fig:life}).}
\begin{ruledtabular}
\begin{tabular}{rllll}
isotope & \multicolumn{1}{l}{$\lambda^\mathrm{(A)}_m$ (s$^{-1}$)} &
          \multicolumn{1}{l}{$\lambda^\mathrm{(A)}_g$ (s$^{-1}$)} &
          $c^\mathrm{(A)}_{m}$ (s$^{-1}$)                         &
          $c^\mathrm{(A)}_{g}$ (s$^{-1}$)                         \\
 \hline\rule[0mm]{0mm}{4mm}
$A=48$ & 0.070(2)& 0.0202(5) & 161(35) & 831(48) \\
$A=47$ & 0.62(3) & 0.01665(6) & ~~~9.8(3) & ~\,33.86(6) \\
\end{tabular}
\end{ruledtabular}
\end{table}

With this assumption the fitted rate constants can be expressed as
\cite{Schmidt1994}
\begin{equation}\label{eq:lambda}
  \lambda_g^\mathrm{(A)} = A_{gl}^\mathrm{(A)}\mathrm{~~~and~~~}
  \lambda_m^\mathrm{(A)} = A_r^\mathrm{(A)}+
  A_\mathrm{DC}+A_{ml}^\mathrm{(A)}
\end{equation}
where $A_r$ is the radiative decay rate of the $^3P_0$ state and
$A_\mathrm{DC}$ is the decay rate of this state due to dielectronic
capture (DC) into the $(2s\,2p\;^1P_1)\,11d$ and $(2s\,2p\;^3P_2)\,26l$
doubly excited states (Fig.\ \ref{fig:DR}). Loss of $^3P_0$ ions occurs
if these states decay either via photon emission (in this case DR has
occurred) or via autoionization to the $2s^2\;^1S_0$ ground state. Since
DC and the subsequent relaxation processes are not significantly
influenced by hyperfine effects the rate $A_\mathrm{DC}$ is the same for
both isotopes. The rates $A_{gl}$ and $A_{ml}$ describe the loss of
$^1S_0$ ground state and of $^3P_0$ metastable ions, respectively, from
the storage ring. The most important processes that lead to the loss of
ions from the storage ring are collisions with residual-gas particles and
electron-ion recombination in the electron cooler.

The loss rate $A_{gl}$ is different for the two isotopes because the
$^{47}$Ti$^{18+}$ and $^{48}$Ti$^{18+}$ ions are stored with different
velocities $v$ and because of the velocity dependence of the relevant
cross sections $\sigma$. With $\sigma \propto v^{-x}$ it follows that
$A_{gl} \propto v\sigma \propto v^{1-x}$. The exponent $x$ can be
determined from $A_{gl}^\mathrm{(48)}/A_{gl}^\mathrm{(47)} =
[\beta^\mathrm{(48)}/\beta^\mathrm{(47)}]^{1-x}$ yielding $x=10(1)$. This
value is consistent with the empirical $v^{-9.6}$ scaling
\cite{Schlachter1983} of the cross section for charge capture during
collisions of highly charged ions with neutral residual gas particles.

The ionization energies of the $^1S_0$ ground state and the $^3P_0$
metastable state, 1346~eV and 1310~eV, respectively
\cite{Ralchenko2005a}, differ by less than 3\%. Therefore, it can
safely be assumed that the loss rates for both states are
approximately equal, i.\,e., that $A_{ml}=A_{gl}$. This assumption
does not lead to any serious consequences. As will be shown below,
the final result for the HFI transition rate changes
only insignificantly when, e.\,g., $A_{ml}=2 A_{gl}$ is assumed.

With the assumption of $A_{ml} = A_{gl}$ Eq.\ \ref{eq:lambda} can be
solved for $A_r$ yielding $A_r^\mathrm{(A)} =
\lambda_m^\mathrm{(A)}-\lambda_g^\mathrm{(A)}-A_\mathrm{DC}$. Since
$A_r^\mathrm{(48)} = 0$ the DC rate is
$A_\mathrm{DC}=\lambda_m^\mathrm{(48)}-\lambda_g^\mathrm{(48)}=0.050(2)$~s$^{-1}$.
Finally, the HFI transition rate of $^{47}$Ti$^{18+}$ state is calculated
as
\begin{eqnarray}
 A_\mathrm{HFI} &=& \gamma^{(47)}A_r^\mathrm{(47)}
 =
 \gamma^{(47)}[\lambda_m^\mathrm{(47)}-\lambda_g^\mathrm{(47)}-A_\mathrm{DC}]\\
 &=& \gamma^{(47)}[\lambda_m^\mathrm{(47)}-\lambda_g^\mathrm{(47)}-
     \lambda_m^\mathrm{(48)}+\lambda_g^\mathrm{(48)}]\nonumber
\end{eqnarray}
where the relativistic factor
$\gamma^{(47)}=[1-(\beta^{(47)})^2]^{-1/2}=1.00531(1)$ occurs because of
the transformation into the ion's frame of reference. With the values for
$\lambda_{m,g}^\mathrm{(A)}$ from Tab.\ \ref{tab:fit} one obtains
$A_\mathrm{HFI} = 0.56(3)$~s$^{-1}$. With the assumption of $A_{ml}=2
A_{gl}$ (see above) the result would change to $A_\mathrm{HFI} =
0.54(3)$~s$^{-1}$. This change is within the experimental uncertainty
which is mainly determined by the statistical uncertainty of
$\lambda_m^\mathrm{(47)}$ (Tab.\ \ref{tab:fit}). It shows the relatively
low sensitivity of the experimental radiative rate on the difference
between $A_{ml}$ and $A_{gl}$.

The remaining issue to be discussed is the possible quenching of the
$^3P_0$ state in the magnetic fields of the storage ring magnets
\cite{Mannervik1996a} via the $B$-field induced mixing of the $^3P_0$
state with the $^3P_1$ state. The magnitude of the mixing coefficient is
of the order of $\mu_B B/\Delta E$ where $\mu_B$ is the Bohr magneton and
$\Delta E$ is the $^3P_0 - {^3P_1}$ energy splitting. For the present
experiment it is estimated that the $^3P_0\to ^1S_0$ transition rate  by
$B$-field induced mixing is more than two orders of magnitude smaller
than $A_\mathrm{HFI}$. Therefore, it can safely be neglected.

The present experimental value $A_\mathrm{HFI} = 0.56(3)$~s$^{-1}$ for
the hyperfine induced $^3P_0 \to {^1S_0}$ transition rate in Be-like
$^{47}$Ti$^{18+}$ with $I=5/2$ is 57\% larger than the theoretical value
of 0.3556~s$^{-1}$ \cite{Marques1993}. It has been shown that calculated
values of $A_\mathrm{HFI}$ are very sensitive to electron-electron
correlation \cite{Brage1998a}. If correlation is treated more thoroughly,
HFI transition rates larger by factors of up to 4 \cite{Brage1998a} (see
above) are obtained as compared with a less extensive treatment
\cite{Marques1993}.  Therefore, the present discrepancy between
experiment and theory is ascribed to a partial neglect of important
correlation effects in the theoretical calculation.

In summary, in this work an experimental value from a laboratory
measurement is presented for the very low hyperfine induced $^3P_0 \to
{^1S_0}$ transition rate in Be-like Ti$^{18+}$. It is almost an order of
magnitude more precise than the only previous experimental value for
isoelectronic N$^{3+}$ \cite{Brage2002a} that was obtained from
astrophysical observations and modeling. The present value for the HFI
transition rate exceeds the only presently available theoretical result
\cite{Marques1993} by 57\%. This difference is attributed to electron
correlation effects that were included only  approximately in the
theoretical calculation. An essential feature of the present experimental
method is the comparison of measured results from different isotopes with
zero and nonzero nuclear spin. The method is readily applicable to a wide
range of ions and has the potential for yielding even more accurate
results.

The authors thank M. Schnell for a helpful discussion and gratefully acknowledge the excellent
support by the MPI-K accelerator and TSR crews. This work was supported in part by the German federal
research-funding agency DFG under contract no.\ Schi~378/5.

\end{document}